\documentclass[english,doc]{article}
\usepackage[T1]{fontenc}
\usepackage[latin9]{inputenc}
\usepackage{mathrsfs}
\usepackage{amsmath}
\usepackage{graphicx}
\usepackage[numbers]{natbib}

\makeatletter

\let\SF@@footnote\footnote
\def\footnote{\ifx\protect\@typeset@protect
    \expandafter\SF@@footnote
  \else
    \expandafter\SF@gobble@opt
  \fi
}
\expandafter\def\csname SF@gobble@opt \endcsname{\@ifnextchar[%]
  \SF@gobble@twobracket
  \@gobble
}
\edef\SF@gobble@opt{\noexpand\protect
  \expandafter\noexpand\csname SF@gobble@opt \endcsname}
\def\SF@gobble@twobracket[#1]#2{}
\providecommand{\tabularnewline}{\\}

\usepackage{babel}

\makeatother

\usepackage{babel}
\begin{document}

\title{Hilbert Space Multi-dimensional Modeling}

\author{Jerome R Busemeyer \\
 %EndAName
Indiana University \and Zheng Wang \\
 %EndAName
Ohio State University }
\maketitle
\begin{abstract}
This article presents general procedures for constructing, estimating,
and testing Hilbert space multi-dimensional (HSM) models, which are
based on quantum probability theory. HSM models can be applied to
collections of $K$ different contingency tables obtained from a set
of $p$ variables that are measured under different contexts. A context
is defined by the measurement of a subset of the $p$ variables that
are used to form a table. HSM models provide a representation of the
collection of $K$ tables in a low dimensional vector space, even
when no single joint probability distribution across the $p$ variables
exists. HSM models produce parameter estimates that provide a simple
and informative interpretation of the complex collection of tables.
Comparisons of HSM model fits with Bayes net model fits are reported
for a new large experiment, demonstrating the viability of this new
model. We conclude that the model is broadly applicable to social
and behavioral science data sets.
\end{abstract}
When large data sets are collected from different contexts, often
they can be summarized by collections of contingency tables or cross-tabulation
tables. Suppose there are $p$ different variables $\left(Y_{1},\ldots,Y_{p}\right)$
that can be used to measure objects, or events, or people. It may
not be possible to measure all $p$ variables at once, and instead,
only a subset of variables ($Y_{k_{1}},\ldots,Y_{k_{s}}$), $s<p,$
can be measured at once. Each subset forms a context $k$ of measurement.
More than one context can be collected, which forms a collection of
$K$ data tables $(T_{1},\ldots,T_{k},\ldots T_{K})$, each collected
under a different context $k$. Each table $T_{k}$ is a joint relative
frequency, or contingency, table based on a subset of variables. 

For example, a research problem could involve three variables $(Y_{1},Y_{2},Y_{3})$,
but some tables might include only some subset of the three variables.
One context might involve the measurement of a single variable $Y_{1}$
that has 5 values to form a $1-way$ frequency table $T_{1}$ composed
of 5 frequencies. Another context could be used to form another $5\times3$
table $T_{2}$, composed of joint frequencies for two variables $(Y_{1},Y_{2})$.
A third context could form another $3\times4$ table $T_{3}$ containing
variables $(Y_{2},Y_{3})$, and fourth could form a $5\times4$ table
containing variables $(Y_{1},Y_{3}).$ 

A critical problem arises: How to integrate and synthesize these $K$
different tables into a compressed, coherent, and interpretable representation?
It is common to apply categorical data analysis \citep{AgrestiCatDatBook}
to a single $p-way$ table (e.g., a single $5\times3\times4$ table).
However, the problem is different here because there are a collection
of $K$ tables of varying dimensions rather than a single $p-$dimensional
table. 

A common solution is to assume that the $K$ tables are generated
from a \emph{single} latent $p-way$ joint distribution, and then
try to reproduce the frequencies in the $K$ different tables by marginalizing
across variables in the $p-way$ table. Often Bayesian causal networks
are used to reduce the number of latent probability parameters by
imposing conditional independence assumptions \citep{darwiche2009modeling,Anderson2002}.
Unfortunately, however, in many cases, no such $p-way$ joint distribution
exists that can reproduce the observed tables! This occurs when the
data tables violate consistency constraints required by classical
(Kolmogorov) probability theory upon which Bayes nets are built; in
this case, no Bayesian network representation composed of the $p$-variables
can even be formed. For example, the tables may have inconsistent
marginal probabilities for a variable $Y_{1}$, or the probabilities
assigned to two sequential measures $(Y_{1},Y_{2})$ may not be commutative.
In the following sections, we give concrete examples of the various
types of possible joint probability violations.

Hilbert space multidimensional (hereafter, denoted HSM) modeling is
based on quantum probability theory \citep{GudderQP,Pitowski_1989,suppes1961probability}.
It provides a promising new solution to these problems faced by complex
data by constructing a model that has (a) a single finite state vector
that lies within a low dimensional vector space, and (b) by forming
a set of measurement operators that represent the $p$ measurements.
In this way, we can achieve a compressed, coherent, and interpretable
representation of the $p$ variables that generate the complex collection
of $K$ tables, even when no standard $p$-way joint distribution
exists. In a Hilbert space model, the state vector represents respondents\textquoteright{}
initial tendencies to select responses to each of the $p$ measurements;
the measurement operators describe the inter-relations between the
$p$ measurements (independent of the initial state of the respondents).
\footnote{Technically, a Hilbert space is a complete inner product vector space
defined on a complex field. Our vector spaces are finite, and so they
are always complete. } 

HSM models are similar to traditional multidimensional scaling (MDS)
models \citep{young2013multidimensional}, but also different from
them in important aspects. Like traditional MDS models, HSM models
are based on similarity relations between entities located within
a vector space. However, traditional MDS models define the similarity
relations by inner products between vectors in a real vector space,
whereas HSM models define similarity relations by projections onto
subspaces of a complex vector space. Also, MDS models are designed
to account for a single $2-way$ similarity matrix, whereas HSM models
can be applied to multiple similarity matrices (e.g., when the similarity
relation is asymmetric, see \citealt{PothosBusemeyerSim}).

The article is organized as follows. First, we briefly justify our
extension and application of quantum probability theory to study social
and behavioral sciences. Second, we provide an artificial data example
that illustrates how consistency requirements of a single $p-way$
joint distribution can be violated. Third, we describe the general
procedures for building HSM models. Fourth, we provide a concrete
application to the artificial data. Fifth, we present an application
of the principles to real data obtained from evaluations of public
service announcements. Finally, we finish with a summary of the new
contributions made by HSM models.

\section{Why Apply Quantum Theory to Social and Behavioral Sciences?}

Classical probability theory evolved over several centuries, beginning
in the 18th century with contributions by Pascal and Laplace. However,
an axiomatic foundation for classical probability theory did not exist
until \citet{Kolmogorov} provided one. Much of the theory was initially
motivated by problems arising in physics, and later applications appeared
in economics, engineering, insurance, statistics, etc. Classical probability
theory is founded on the premise that events are represented as subsets
of a larger set called the sample space. The adoption of subsets as
the basis for describing events entails a logic\textemdash the logic
of subsets\textemdash which is equivalent to Boolean logic (more generally,
a sigma algebra of events). Boolean logic includes some strict laws,
such as the closure property that if A, B are events then $A\cap B$
is an event, and the axiom that events are commutative, $(A\cap B)=(B\cap A)$,
and distributive, $A\cap(B\cup C)=(A\cap B)\cup(A\cap C)$. Social
and behavioral scientists are generally trained to accept these axioms
(explicitly or implicitly), and consequently most of us consider the
theory as the only way to think about events and probabilities. How
could there be other ways? 

Looking back into the history, scientists were faced with similar
questions, such as with Euclidean geometry. How could there be any
other geometry other than Euclidean? Nevertheless, we now have many
applications of non-Euclidean geometry. Could this happen with probability
theory too? Quantum mechanics was invented by a brilliant group of
physicists in the 1920\textquoteright s in response to physical phenomena
that seemed paradoxical from a classical physics perspective. This
theory has revolutionized our world by giving us transistors, lasers,
a foundation for chemistry, and many other accomplishments. Interestingly,
though not at first realizing it, these physicists invented an entirely
new theory of probability. It was not clear that they invented a new
probability theory until an axiomatic foundation was provided by \citet{Dirac_book}
and \citet{VonNeumann_1932}. Quantum theory is founded on the premise
that events are represented as subspaces of a vector space (called
a Hilbert space, hence the name of our model). The adoption of subspaces
as the basis for describing events entails a new logic\textemdash the
logic of subspaces\textemdash which relaxes some of the axioms of
Boolean logic. In particular, this logic does not entail that events
are always commutative and distributive, and the closure property
does not always hold. 

It turns out that quantum probability theory is not only useful for
explaining physical phenomena, but it also provides useful new tools
to model human behavior \citep{PothosBuseBBS}. Notice we are not
proposing that the brain is some kind of quantum computer, and instead,
we are only using the mathematical principles of quantum theory to
account for human behavior. We have found many cases where quantum
probability theory provides a better account of human judgment and
decisions than classical probability theory \citep{BruzaWangBuseTiCS2015}.
In particular, human judgments are not commutative, and order effects
are pervasive \citep{WangPNAS2014,TruebloodQuantInf2010}; human decisions
often violate the law of total probability that follows from the distributive
axiom \citep{PothosBusemeyer_2009,WangBuse_Cognition_2016,KvamPNAS2015};
quantum theory provides a coherent account of many different types
of probability judgment errors \citep{BusemeyerQP2011,Aerts_JMP_2009}
as well as violations of rational decision making \citep{LaMura_2009,YukalovSornette2010,asano2016quantum}.
Quantum theory provides a natural account of asymmetry in similarity
judgments \citet{PothosBusemeyerSim}. We could go on with more examples,
but a review of this rapidly growing literature is beyond the purpose
of this article. 

The principles from quantum theory actually resonate with deeply rooted
psychological conceptions \citep{AertsBroekaertGabora2011}. First,
consider the enigmatic quantum principle of superposition\textemdash it
captures the intuitive feelings of conflict, ambiguity, or uncertainty.
A superposition state is maintained across potential choices until
a decision must be reached, at which point the state collapses to
a specific choice \citep{LambertMogiliansky2009}. This behavior of
changing from a superposition to a specific decision is similar to
what Bohr called the wave-particle aspects of quantum mechanics. Next,
consider the principle of complementarity\textemdash taking a measurement
of a system constructs rather than records a property of the system,
and the first question sets up a context that changes the answer to
the next question, thus answering a question disturbs the answers
to subsequent questions and the order of questions is important \citep{WangBuseAtmPothos2013}.
In quantum physics, order-dependent measurements are said to be non-commutative
and quantum theory was especially designed for these types of non-commutative
measures. Finally, consider the unique quantum concept of entanglement\textemdash the
event $A\cap B$ may be observed, and another event $C\cap D$ maybe
observed, but the event $A\cap B\cap C\cap D$ may not even exist,
violating closure. Quantum probabilities based on an entangled state
provides a basis for explaining these types of non-classical systems
\citep{bruza2015probabilistic,Aerts&Gabora_2005}.\footnote{Reviewers often argue that although the micro world is quantum, the
macro world that we observe is classical, and so why would nature
evolve a non-commutative human reasoning system? This confuses an
important point. We are comparing classical versus quantum probability
models of observed (epistemic) phenomena. We are not comparing classical
versus physical models of an unobserved (ontological) macro world.
Even classical physical models of the world can produce observed probabilities
that are non-commutative. The latter can happen when only coarse epistemic
measurements of the underlying ontic physical states are available
\citep{Graben&Atmanspacher2006}. }

\section{An Artificial Example}

In this section we present an artificial example that serves to illustrate
several ways that a joint probability model can fail. Suppose that
four variables are measured, labeled A,H,I, and U. For example, suppose
queries are made from a large social media source on political candidates
concerning Attractiveness, Honesty, Intelligence, and Unusualness.
As another example, suppose queries are made from a large medical
record source on patient symptoms concerning Anxiety, Hyperactivity,
Irritation, and Unruliness. As a third example, suppose queries are
made from a large consumer choice source about food products concerning
whether the product is Appetizing, Healthy, Interesting, and Unfamiliar.
It is difficult or impossible to obtain ratings from individuals on
all four attributes simultaneously. Suppose that only pairs of attributes
are queried at a time, for example, the pair A,I and the pair A,H.
Each pair provides a context for answering the questions.

For simplicity, suppose each query is answered with a yes (Y) or no
(N) answer. Thus a pair of yes-no answers to a pair of attributes
forms one $2\times2$ table with relative frequencies for pairs of
answers YY,YN,NY,NN. Suppose $8$ contexts are used to form $8$ different
$2\times2$ tables as shown in Table 1. For example, the pair of attributes
AI form the context for the second $2\times2$ table. Each cell within
a row is a relative frequency for a $2\times2$ table, and the cells
within a row sum to one. For example, the relative frequency of yes
to attribute A and no to attribute I equals .175. Note that question
ordering may matter so that, for example, the context AH is treated
different from HA. For simplicity, we only included a subset of all
12 possible $2-way$ tables. These $8$ tables are sufficient to make
our points. 
\begin{table}
\caption{\label{tab:ArtificialData}Eight different $2\times2$ tables produced
by yes, no answers to pairs of attributes A,H,I,U. The label YN refers
to yes to the first and no to the second attribute. Each cell within
a row is a relative frequency, and all the cells within a row sum
to one. The order of questions may matter so that the HA table (H
asked before A) may differ from the AH table (A asked before H).}

\centering{}%
\begin{tabular}{ccccc}
Pair & YY & YN & NY & NN\tabularnewline
\hline 
AH & .345 & .101 & .125 & .429\tabularnewline
\hline 
AI & .271 & .175 & .084 & .469\tabularnewline
\hline 
AU & .115 & .331 & .269 & .285\tabularnewline
\hline 
HI & .335 & .035 & .021 & .610\tabularnewline
\hline 
HU & .296 & .073 & .088 & .543\tabularnewline
\hline 
IU & .300 & .055 & .100 & .545\tabularnewline
\hline 
HA & .286 & .083 & .143 & .488\tabularnewline
\hline 
UI & .325 & .059 & .095 & .521\tabularnewline
\hline 
\end{tabular}
\end{table}

\subsection{Does a joint distribution exist?}

The following question can be asked about Table \ref{tab:ArtificialData}:
Does a \emph{single} $4-way$ joint probability distribution exist
that can reproduce Table \ref{tab:ArtificialData}? The $4-way$ joint
probability distribution is defined by $4$ binary random variables
$(A,H,I,U)$ that generate $16$ latent joint probabilities that sum
to one: $\pi(A=w\cap H=x\cap I=y\cap U=z),$ where, for example, $A$
is a random variable with values $w=1$ for yes and $w=0$ for no,
and similar definitions hold for the other three random variables.
For example, the relative frequency, $p(YN|AI),$ of YN in the context
of the pair AI is predicted by the marginal $\pi(A=1,I=0)=\sum_{x}\sum_{z}\pi(A=1\cap H=x\cap I=0\cap U=z).$
Note that this $4-way$ joint distribution is completely general (non-parametric),
because no conditional independence or parametric distribution assumptions
are imposed. 

The answer to the above question is negative: There is no single $4-way$
joint distribution that can reproduce Table \ref{tab:ArtificialData}.
First of all, the $4-way$ distribution requires the marginal distribution
of a single random variable to be invariant across contexts. This
requirement fails. For example, the marginal probability of yes to
random variable $I$ is not invariant: $p(YY|IU)+p(YN|IU)=.355$ which
differs from $p(YY|UI)+p(NY|UI)=.420$. Table \ref{tab:ArtificialData}
contains other examples of violations of marginal invariance, depending
on whether the attribute appeared first or second. The latter fact
brings up a second problem: the order that questions are asked changes
the $2-way$ distributions for some pairs. For example, the distribution
for the context AH is not the same as the distribution for the context
HA, and an order effect also occurs for the two contexts UI and IU.
Order effects violate the commutative property required by the joint
probability model: in particular, $\pi(A=w\cap H=x)=\pi(H=x\cap A=w),$
and $\pi(I=y\cap U=z)=\pi(U=z\cap I=y),$

It is interesting to notice that in this example, both marginal invariance
and commutativity (no order effects) are satisfied by the four contexts
AI, AU, HI, HU. Suppose we restrict our question to only these four
tables, can a $4-way$ joint distribution reproduce these $4$ tables?
Surprisingly, the answer is still negative. These $4$ tables violate
a consistency requirement of a single $p-way$ joint distribution,
called the Clauser, Horne, Shimony, and Holt (CHSH) inequality \citep[for applications in psychology, see][]{bruza2015probabilistic,DzhafarovKujala2012}.\footnote{The CHSH inequality is closely related to the Bell inequality, and
the latter was derived for the Bohm experiment using a pair of entangled
spin $\frac{1}{2}$ photons, which was used to test the famous Einstein
Podolsky Rosen (EPR) paradox. There are different ways to derive the
CHSH inequality, and we follow the derivation by \citet{Fine:1982}.} The CHSH inequality implies the following restriction on the joint
probabilities required by the $4-way$ joint probability model: $-1\leq CHSH\leq2$,
where

\begin{equation}
CHSH=E(A\cdot I)+E(H\cdot I)+E(H\cdot U)-E(A\cdot U),\label{eq:CHSH}
\end{equation}
and, for example, $E(A\cdot I)=\pi(A=1\cap I=1)$ is the expectation
of the product of the two random variables, $A,I.$ If we set $\pi(A=1\cap I=1)=p(YY|AI),$
$\pi(H=1\cap I=1)=p(YY|HI),$ $\pi(H=1\cap U=1)=p(YY|HU),$ $\pi(A=1\cap U=1)=p(YY|AU)$
from Table \ref{tab:ArtificialData}, then the CHSH value computed
from Table \ref{tab:ArtificialData} equals $CHSH=2.25$, which exceeds
the bound required by the $4-way$ joint probability model. The CHSH
is only one of a number of constraints that are required for a single
joint distribution to reproduce a collection of contingency tables.
Another inequality applies to $3-way$ joint distributions \citep{Suppes:Zanotti:1981,LeggettGarg1985}.
\citet{DzhafarovKujala2012} derive and provide a general summary
of all these linear constraints required for a single joint distribution
to reproduce a collection of contingency tables.

\subsection{Non-parametric statistical tests of the joint distribution model}

Suppose the data in Table \ref{tab:ArtificialData} are based on a
sample of $N=100$ independent observations for each $2\times2$ table.
Then it is unclear whether the violations of the $4-way$ joint probability
distribution, described above, are statistically significant. To address
this issue, we propose the following general method: We compare the
$4-way$ joint probability model to a saturated model.\footnote{This is what \citealt{DzhafKajulaPlos12013} call the \emph{all possible
couplings} model.} The saturated model simply assumes that we have $8$ independent
$2-way$ tables, and each table has $4$ probabilities that sum to
one. The $4-way$ joint probability model has $15$ free parameters,
because the 16 joint probabilities are constrained to sum to one.
The saturated model has $8\times3=24$ parameters, because the probabilities
sum to one within each table. The $4-way$ joint probability model
is nested within the saturated model, and the difference in number
of parameters equals $df=24-15=9.$ Maximum likelihood methods can
be used to estimate the parameters of each model, and $G^{2}=-2\times loglikelihood$
can be determined for each model. Then a likelihood ratio (i.e., chi-square
difference) test can be used to compare models. Using this method
with $N=100$ observations per table produces a chi-square difference
equal to $G_{diff}^{2}=G_{joint}^{2}-G_{sat}^{2}=18.04$, which is
a statistically significant difference with $p=.031$. Therefore,
using this classical statistical test, the joint probability model
is rejected. Note that this is a non-parametric test that requires
no conditional independence or parametric distribution assumptions.

The above non-parametric method for testing a single $4-way$ joint
distribution model can be generalized and applied to $p-way$ joint
distributions as long as there is a sufficient number of tables that
allow the saturated model to have more parameters than the joint distribution
model. For example, if only the four $2\times2$ tables (AI,AU,HI,HU)
are included in the design, then the saturated model has only $4\times3=12$
parameters, which is fewer than the $4-way$ joint distribution model
(see, e.g., \citealt{bruza2015probabilistic}).\footnote{Nevertheless, $G_{diff}^{2}=2.56$ after fitting the $4-way$ model
to Table \ref{tab:ArtificialData}, which reflects the violation of
the CHSH inequality. } However, if four $1-way$ tables, produced by measuring each attribute
alone, are included into the design to form a collection of 8 tables
(A,H,I,U, AI,AU,HI, HU), then the saturated model has $16$ parameters,
which leaves $df=1$ for testing the joint probability model. It is
worth nothing that the number of parameters in the joint probability
model grows exponentially with the number of random variables, which
makes it necessary to impose restrictions (e.g., using Bayesian networks
or parametric distributional assumptions) to form testable models.

Including $1-way$ tables into the design provides direct tests of
the marginal invariance assumption of the joint distribution model.
For example, suppose attributes A,B are measured by binary choices,
and the design included the three tables (A,B,AB). This simple design
provides $2$ $df's$ for testing the joint distribution model (see,
e.g., \citealt{WangBuse_Cognition_2016}). Alternatively, including
different orders of presentation provides direct tests of commutativity.
For example, if attributes A,B are measured on $9$-point scales,
and the design includes both $2-way$ tables (AB,BA), then $df=80$
for testing the joint probability model (see, e.g., \citealt{wang2016comparing}). 

On the one hand, an advantage of this non-parametric statistical test
of the joint distribution model is that it tests all of the constraints
imposed by the joint distribution model (including marginal invariance,
absence of order effects, CHSH inequality, and others) with a single
test. On the other hand, it does not isolate the particular property
that is violated. We have developed more specific log likelihood statistical
tests that are designed to test a particular property (e.g., a test
of order effects versus a test of marginal invariance), but these
additional tests are not described in detail here. 

\subsection{Previous research testing a joint distribution with multiple tables}

The commutative property has been tested by using the pair of tables
(AB,BA) that vary question order. It has long been known that question
order effects commonly occur with human judgments \citep{Tourangeau2000,SchumanPresser1981}.
Recently, quantum models have provided good accounts for these effects
\citep{WangBusemeyerOrderEffect,WangPNAS2014,wang2016comparing}.
The marginal invariance assumption has been tested by using a design
with two tables (A,BA). These are also called tests of the law of
total probability, or tests for interference effects. Several experiments
have been conducted that demonstrate violations of marginal invariance
\citep{BuseWangMog_2009,DiNunzioBruzaSitbon2014,Conte2009,KvamPNAS2015,wang2016exploration,WangBuse_Cognition_2016}.
A number of experiments have been conducted to test the CHSH or similar
inequalities required by a single joint distribution applied to a
collection of several $2\times2$ tables \citep{bruza2015probabilistic,AertsGaboraSozzo2013,asano2014violation,gronchi2016quantum}.
Although violations of the required inequalities were reported in
the experiments testing the CHSH inequality, they were confounded
with violations of marginal invariance \citep{dzhafarov2016there}.
It remains to be found out whether or not human judgments produce
violations of the CHSH inequality in the absence of violations of
order effects and marginal invariance \citep{dzhafarov2016there}. 

\section{Multidimensional Hilbert Space Modeling}

\subsection{Basics of quantum probability theory}

It is helpful to introduce quantum probability theory by comparing
it with classical probability theory.\footnote{See \citet{BusemeyerBruza2012,HavenKhrenBook,Khrennikov2010,Rijsbergen:QM04}
for introductions to quantum probability theory written for social
and behavioral sciences. } Although both classical and quantum theories are applicable to infinite
spaces, for simplicity, we limit this presentation to finite spaces. 

Suppose we have $p$ variables $(Y_{i},i=1,\cdots,p)$ and each variable,
such as $Y_{i}$, produces one of a finite set of $n_{i}$ values
when measured. In classical theory, variable $Y_{i}$ is called a
random variable, and in quantum theory, $Y_{i}$ is called an observable.
The measurement outcome generated by measuring one of the $p$ variables
produces an event. For example, if variable $Y_{1}$ is measured and
it produces the value $y_{i}$, then we observe the event $(Y_{1}=y_{i})$. 

Classical theory begins with a universal set $\Omega$ containing
all events, which is called the sample space; and quantum theory replaces
this with a vector space $\mathscr{H}$ containing all events, which
is called the Hilbert space. Classical theory defines an event $A$
as a \emph{subset} of the sample space, whereas quantum theory defines
an event $A$ as a \emph{subspace} of the Hilbert space. Each subspace,
such as $A$, corresponds to a projector, denoted $P_{A}$ for subspace
$A,$ which projects vectors into the subspace. The change from subsets
to subspaces is where the logic of events differs between the two
theories. 

Classical theory assumes closure: If $A\in\Omega$ is an event, and
$B\in\Omega$ is another event, then $A\cap B\in\Omega$ is also an
event in the sample space. By definition of intersection, the classical
event $A\cap B$ is commutative $A\cap B=B\cap A$. In quantum theory,
the events $A\in\mathscr{H}$, $B\in\mathscr{H}$ may not be commutative,
and if they are not, then the conjunction does not exist, and closure
does not hold. Instead, quantum theory uses the more general concept
of a sequence of events.

In quantum theory, a sequence of events, such as $A$ and then $B$,
denoted $AB$, is represented by the sequence of projectors $P_{B}P_{A}$.
If the projectors commute, $P_{A}P_{B}=P_{B}P_{A}$, then the product
of the two projectors is a projector corresponding to the subspace
$A\cap B$, that is, $P_{B}P_{A}=P(A\cap B)$; and the events \emph{A}
and \emph{B} are said to be \emph{compatible}. However, if the two
projectors do not commute, $P_{B}P_{A}\neq P_{A}P_{B}$, then neither
their product is a projector, and the events are \emph{incompatible}. 

Classical theory defines a set function $p$ that assigns probabilities
to events, which is required to be an additive measure: $p(A)\geq0,$
$p(\Omega)=1,$ and if $A\cap B=\oslash$, then $p(A\cup B)=p(A)+p(B)$.
Quantum theory uses a unit length state vector, denoted $\left|\psi\right\rangle \in\mathscr{H}$,
to assign probabilities to events as follows:\footnote{A more general approach uses what is called a density operator rather
than a pure state vector, but to keep ideas simple, we use the latter.} 
\begin{equation}
p(A)=\left\Vert P_{A}\left|\psi\right\rangle \right\Vert ^{2},\label{eq:QuantumProb}
\end{equation}
Quantum probabilities also satisfy an additive measure: $p(A)\geq0,$
$p(\mathscr{H})=1,$ and if $P_{A}P_{B}=0,$ then $p(A\vee B)=p(A)+p(B)$.
In fact, Equation \ref{eq:QuantumProb} is the unique way to assign
probabilities to subspaces that form an additive measure for dimensions
greater than 2 \citep{Gleason1957}.

According to classical theory, if an event $A$ is an observed fact,
then the conditional probability of event $B$ is defined as 
\[
p(B|A)=\frac{p(A\cap B)}{p(A)},
\]
and so the joint probability of $A\cap B$ equals $p(A\cap B)=p(A)\cdot p(B|A).$
The corresponding definition in quantum theory is 
\[
p(B|A)=\frac{\left\Vert P_{B}P_{A}\left|\psi\right\rangle \right\Vert ^{2}}{p(A)},
\]
and so the probability of the sequence $AB$ equals $p(AB)=p(A)\cdot p(B|A)=\left\Vert P_{B}P_{A}\left|\psi\right\rangle \right\Vert ^{2}.$
The commutative property of classical probability requires that $p(A)\cdot p(B|A)=p(B)\cdot p(A|B)$,
but this commutative property does not hold for quantum theory so
that $p(A)\cdot p(B|A)\neq p(B)\cdot p(A|B)$ occurs when events are
incompatible. 

Extensions to sequences with more than two events follows the same
principles for both classical and quantum theories. The probability
of the joint event $\left(A\cap B\right)\cap C$ equals $p(\left(A\cap B\right)\cap C)$
for classical theory, and the probability of the sequence $(AB)C$
equals $\left\Vert P_{C}\left(P_{B}P_{A}\right)\left|\psi\right\rangle \right\Vert ^{2}$
for quantum theory. 

\subsection{Building Projectors}

This section describes a general way to construct the projectors for
events in the Hilbert space, and to formally describe the conditions
that produce incompatibility. This section is somewhat abstract and
technical, and a concrete application is provided in the next section
where we build a simple model for Table \ref{tab:ArtificialData}.
In the following, $\left|V\right\rangle $ denotes a vector in the
Hilbert space, $\left\langle V|W\right\rangle $ denotes an inner
product, $\left|V\right\rangle \left\langle V\right|$ denotes an
outer product, and $P^{\dagger}$ denotes a Hermitian transpose.

In general, a projector, denoted $P$, operating in an $N-$dimensional
Hilbert space $\mathscr{H}$, is defined by the two properties $P=P^{\dagger}=P^{2}$.
By the first property, $P$ is Hermitian, and so it can be decomposed
into $N$ orthonormal eigenvectors; by the second property, $P$ has
only two eigenvalues, which are simply $(0,1)$ \citep{Halmos1993}.
Define $\left|V_{j}\right\rangle ,j=1,\cdots,N$ as the set of $N$
orthonormal eigenvectors of $P$. The projector $P$ can be expressed
in terms of the eigenvectors as follows
\begin{equation}
P=\sum_{j}\lambda_{j}\left|V_{j}\right\rangle \left\langle V_{j}\right|,\label{eq:eigendecomp}
\end{equation}
where the outer product, $\left|V_{j}\right\rangle \left\langle V_{j}\right|$,
is the projector that projects into the ray spanned by eigenvector
$\left|V_{j}\right\rangle $, and $\lambda_{j}=1$ if $\left|V_{j}\right\rangle $
corresponds to an eigenvalue of 1, and $\lambda_{j}=0$ if $\left|V_{j}\right\rangle $
corresponds to an eigenvalue of $0.$ These $N$ eigenvectors form
an orthonormal basis that spans the Hilbert space. Every vector, such
as $\left|\psi\right\rangle \in\mathscr{H}$ can be expressed as a
linear combination of these basis (eigen) vectors
\begin{equation}
\left|\phi\right\rangle =\sum_{j}^{N}\phi_{j}\cdot\left|V_{j}\right\rangle \label{eq:Psy}
\end{equation}

If two projectors, $P_{A},P_{B}$ share all of the same eigenvectors,
then they commute \citep{Halmos1993}. In other words, two events
$A,B$ are compatible if they are described in terms of the same basis.
If the two projectors do not share all of the same eigenvectors, then
they do not commute, and the events $A,B$ are described by two different
bases. They are incompatible, and must be evaluated sequentially,
because one needs to change from one basis to evaluate the first event,
to another basis to evaluate the second event, making them incompatible. 

Define $\left|V_{j}\right\rangle ,j=1,\cdots,N$ as the basis used
to describe event $A,$ and define $\left|W_{j}\right\rangle ,j=1,\cdots,N$
as the basis used to describe event $B.$ We can change from one basis
to another by a unitary transformation (a ``rotation'' in Hilbert
space)
\begin{equation}
\left|W_{j}\right\rangle =U\left|V_{j}\right\rangle ,j=1,\cdots,N,\label{eq:Unitary}
\end{equation}
where $U$ is defined by $U^{\dagger}U=I$, that is, $U$ is an isometric
transformation that preserves inner products \citep{Halmos1993}.
Therefore, the projector for event $B$ can be re-expressed in terms
of the event $A$ basis $\left|V_{j}\right\rangle ,j=1,\cdots,N$
as follows
\begin{eqnarray}
P_{B} & = & \sum_{j}^{N}\lambda_{j}\left|W_{j}\right\rangle \left\langle W_{j}\right|\nonumber \\
 & = & U\left(\sum\lambda_{j}\left|V_{j}\right\rangle \left\langle V_{j}\right|\right)U^{\dagger}.\label{eq:ProjB}
\end{eqnarray}
According to Equation \ref{eq:Unitary}, the unitary transformation
$U$ represents the transitions from state $\left|W_{i}\right\rangle $
to state $\left|V_{j}\right\rangle $ by the inner product $\left\langle V_{j}|W_{i}\right\rangle $.

So far, we have presented a general method for building the projectors
by defining a basis for the vector space and by transforming from
one basis to another using unitary transformation. Then the next question
is how to build the unitary transformation? In general, any unitary
transformation can be built from a Hermitian operator $H$ as follows
(Halmos, 1993):
\begin{equation}
U=exp(-i\cdot H).\label{eq:buildunitary}
\end{equation}
The right hand side is exponential function of the Hermitian operator
$H$ (see appendix for details).

In summary, the HSM program selects a Hermitian operator $H$ for
Equation \ref{eq:buildunitary}, and then uses the Hermitian operator
to build the unitary operator $U$ which provides the relation between
projectors $P_{A}$ and $P_{B}$ for incompatible events. The beauty
of using a vector space is that it provides an infinite number of
ways to generate incompatible variables by unitary ``rotation,''
and yet remain within the same $N-$dimensional space. This is how
an HSM model maintains a low dimensional representation even when
there are a large number of variables. 

\subsection{Building the Hilbert space}

This section describes how we construct a Hilbert space to represent
the $p$ variables. This construction depends on the compatibility
relations between the variables. For this section, we need to use
the Kronecker (tensor) product between two matrices, denoted as $P\otimes Q$
(see the Appendix for a brief review). 

To begin building the Hilbert space, suppose we measure a single variable,
say $Y_{1}$, that can produce $n_{1}$ values corresponding to the
mutually exclusive and exhaustive set of events $(Y_{1}=y_{i}),i=1,\cdots n_{1}.$
To represent these events in a Hilbert space, we partition the space
into $n_{1}$ orthogonal subspaces. Each subspace, such as $(Y_{1}=y_{i})$,
corresponds to a projector $P(Y_{1}=y_{i}).$ The projectors for all
of the events are pairwise orthogonal, $P(Y_{1}=y_{i})P(Y_{1}=y_{j})=\mathbf{0}$,
and complete, $\sum_{i}P(Y_{1}=y_{i})=\mathbf{I}$ (where $\mathbf{I}$
is the identity that projects onto the entire Hilbert space). These
$n_{1}$ events are all compatible, and the projectors are all commutative,
because they are all orthogonal to each other. Each projector generates
$N_{1}\geq n_{1}$ eigenvectors, and the projectors all share the
same eigenvectors, but with different eigenvalues. These $N_{1}$
eigenvectors provide the basis for spanning a $N_{1}-$dimensional
Hilbert space, $\mathscr{H}_{N_{1}}$. 

Continuing with the case of a single variable represented by the Hilbert
space $\mathscr{H}_{N_{1}}$ , we can express each vector $\left|\phi\right\rangle \mathscr{\in H}_{N_{1}}$
in terms of its coordinates with respect to the eigenvectors of $P(Y_{1}=y_{i})$
by using Equation \ref{eq:Psy}. Using this basis, the coordinate
representation of each projector, say $P(Y_{1}=y_{i})$ is simply
an $N_{1}\times N_{1}$ diagonal matrix, $M_{1}(i)$ with ones located
in the rows corresponding to basis vectors that have an eigenvalue
of one associated with the projector $P(Y_{1}=y_{i})$, and zeros
otherwise. The coordinate representation of $\left|\psi\right\rangle $
with respect to this basis is a $N_{1}\times1$ column matrix $\psi$
with coordinate $\psi_{i}$ in row $i,$ which satisfies $\psi^{\dagger}\psi=1$.
Then the probability distribution over the values of $Y_{1}$ for
$i=1,\cdots,n_{1}$ is given by 
\begin{equation}
\left\Vert P(Y_{1}=y_{i})\cdot\left|\psi\right\rangle \right\Vert ^{2}=\left\Vert M_{1}(i)\cdot\psi\right\Vert ^{2}=\left|\psi_{i}\right|^{2}.
\end{equation}
There is little difference between classical and quantum probability
at this point.

Next suppose we measure two variables, $Y_{1}$ with $n_{1}$ values
and $Y_{2}$ with $n_{2}$ values, with $n_{1}\geq n_{2}$. If these
two variables are compatible, then the joint event $(Y_{1}=y_{i}\cap Y_{2}=y_{j})$
is well defined for all pairs of values. Therefore the Hilbert space
is partitioned into $n_{1}\cdot n_{2}$ orthogonal subspaces. Each
subspace corresponds to a projector $P(Y_{2}=y_{j})P(Y_{1}=y_{i})=P(Y_{1}=y_{i})P(Y_{2}=y_{j})=P(Y_{1}=y_{i}\cap Y_{2}=y_{j})$.
These projectors are pairwise orthogonal and complete, and every pair
of projectors is commutative. Each projector shares $(N_{1}\cdot N_{2})\geq(n_{1}\cdot n_{2})$
eigenvectors, but with different eigenvalues, to span a Hilbert space
$\mathscr{H}_{N_{1}\cdot N_{2}}$. Using this basis, the projector
$P(Y_{1}=y_{i})$ is represented by the Kronecker product $M_{1}(i)\otimes I_{N_{2}}$,
where $I_{N_{2}}$ is an $N_{2}\times N_{2}$ identity matrix. The
projector $P(Y_{2}=y_{j})$ is represented by the matrix Kronecker
product $I_{N_{1}}\otimes M{}_{2}(j)$. Then $P(Y_{2}=y_{j})P(Y_{1}=y_{i})=P(Y_{1}=y_{i}\cap Y_{2}=y_{j})$
is represented by the product $\left(M_{1}(i)\otimes I_{N_{2}}\right)\cdot\left(I_{N_{1}}\otimes M{}_{2}(j)\right)=M_{1}(i)\otimes M{}_{2}(j)$,
which is simply a diagonal matrix with ones located in the rows corresponding
to $(Y_{1}=y_{i}\cap Y_{2}=y_{j})$ and zeros otherwise. The coordinate
representation of $\left|\psi\right\rangle $ with respect to this
basis is a $\left(N_{1}\cdot N_{2}\right)\times1$ column matrix,
$\left(\psi,\psi^{\dagger}\psi=1\right)$, with coordinate $\psi_{ij}$
in row $n_{2}\cdot(i-1)+j$. Then the joint probability for a pair
of values equals 
\begin{equation}
\left\Vert P(Y_{2}=y_{j})P(Y_{1}=y_{i})\left|\psi\right\rangle \right\Vert ^{2}=\left\Vert M_{1}(i)\otimes M{}_{2}(j)\cdot\psi\right\Vert ^{2}=\left|\psi_{ij}\right|^{2}.
\end{equation}
There is still little difference between the classical and quantum
theories at this point. Adding variables increases the dimensionality
of the space, just like it does with a Bayesian model.

Now suppose that variables $Y_{1}$ (with $n_{1}$ values) and $Y_{2}$
(with $n_{2}\leq n_{1}$ values) are incompatible. In this case, we
cannot define the joint occurrence of two events $(Y_{1}=y_{i}\cap Y_{2}=y_{j})$,
and we can only represent a sequence of two single events, e.g., $\left(Y_{1}=y_{i}\right)$
and then $\left(Y_{2}=y_{j}\right)$ by the sequence of projectors
$P(Y_{2}=y_{j})P(Y_{1}=y_{i})$. As before, we define $P(Y_{1}=y_{i})$
as the projector for the event $\left(Y_{1}=y_{i}\right),$ and likewise,
we define $P(Y_{2}=y_{j})$ as projector for the event $\left(Y_{2}=y_{j}\right).$
Both projectors are represented with a Hilbert space, $\mathscr{H}_{N_{1}}$,
of dimension $N_{1}\geq n_{1}$. We can choose to express each vector
$\left|\phi\right\rangle \mathscr{\in H}_{N_{1}}$ in terms of the
coordinates with respect to the eigenvectors of $P(Y_{1}=y_{i})$
by using Equation \ref{eq:Psy}. Using this basis, the coordinate
representation of projector $P(Y_{1}=y_{i})$ is simply an $N_{1}\times N_{1}$
diagonal matrix, $M_{1}(i)$ with ones located in the rows corresponding
to basis vectors that have an eigenvalue of one associated with this
projector, and zeros otherwise. Using Equation \ref{eq:Unitary},
the projector $P(Y_{2}=y_{j})$ can be expressed in terms of the $P(Y_{1}=y_{i})$
basis by a unitary matrix, $U$. Then the matrix representation of
$P(Y_{2}=y_{j})$ is $\left(U\cdot M{}_{1}(j)\cdot U^{\dagger}\right)$.
Finally, the coordinate representation of the state vector $\left|\psi\right\rangle $
with respect to the $Y_{1}$ basis is a $N_{1}\times1$ column matrix
$\psi$. The probability of the sequence of events $\left(Y_{1}=y_{i}\right)$
and then $\left(Y_{2}=y_{j}\right)$ equals
\begin{equation}
\left\Vert P(Y_{2}=y_{j})P(Y_{1}=y_{i})\left|\psi\right\rangle \right\Vert ^{2}=\left\Vert \left(U\cdot M{}_{1}(j)\cdot U^{\dagger}\right)\cdot M{}_{1}(i)\cdot\psi\right\Vert ^{2}.
\end{equation}
This is where a key difference between the classical and quantum theories
occurs. Note that, unlike a Bayesian model, adding variable $Y_{2}$
does not increase the dimensionality of the space.

Finally suppose that we measure three variables, $Y_{1}$ with $n_{1}$
values, $Y_{2}$ with $n_{2}$ values, and $Y_{3}$ with $n_{3}$
values. Suppose $Y_{1}$ is compatible with $Y_{2}$ and $Y_{2}$
is compatible with $Y_{3}$ but $Y_{1}$ is incompatible with $Y_{3}$.
In this case, we can partition the Hilbert space using projectors
$P(Y_{1}=y_{i}\cap Y_{2}=y_{j}),i=1,\cdots n_{1},j=1,\cdots n_{2}$,
which are pairwise orthogonal and complete, and every pair of these
projectors is commutative. Using the eigenvectors of these projectors
as the basis, the projector $P(Y_{1}=y_{i})$ is represented by the
Kronecker product $M_{1}(i)\otimes I_{N_{2}}$, and the projector
$P(Y_{2}=y_{j})$ is represented by the Kronecker product $I_{N_{1}}\otimes M{}_{2}(j)$.
Using a unitary transformation, $U,$ the matrix representation of
the projector $P(Y_{3}=y_{k})$ is given $\left(U\cdot M_{1}(k)\cdot U^{\dagger}\right)\otimes I_{N_{2}}$.
Then, the probability of the two compatible events $\left(Y_{1}=y_{i}\right)$
and$\left(Y_{2}=y_{j}\right)$ equals 

\begin{equation}
\left\Vert P(Y_{2}=y_{j})P(Y_{1}=y_{i})\left|\psi\right\rangle \right\Vert ^{2}=\left\Vert M_{1}(i)\otimes M{}_{2}(j)\cdot\psi\right\Vert ^{2}.
\end{equation}
the probability of the two compatible events $\left(Y_{2}=y_{i}\right)$
and $\left(Y_{3}=y_{j}\right)$ equals 

\begin{equation}
\left\Vert P(Y_{3}=y_{k})P(Y_{2}=y_{i})\left|\psi\right\rangle \right\Vert ^{2}=\left\Vert \left(U\cdot M{}_{1}(k)\cdot U^{\dagger}\right)\otimes M{}_{2}(i)\cdot\psi\right\Vert ^{2},
\end{equation}
and the probability of the sequence of two incompatible events $\left(Y_{1}=y_{i}\right)$
and then $\left(Y_{3}=y_{k}\right)$ equals
\begin{equation}
\left\Vert P(Y_{3}=y_{k})P(Y_{1}=y_{i})\left|\psi\right\rangle \right\Vert ^{2}=\left\Vert \left(U\cdot M{}_{1}(k)\cdot U^{\dagger}\cdot M{}_{1}(i)\right)\otimes I_{N_{2}}\cdot\psi\right\Vert ^{2}.
\end{equation}

The methods described above generalize in a fairly straightforward
manner for more variables. Note that when variables are compatible,
quantum probability theory works like classical probability theory,
and the Hilbert space dimensionality increases exponentially as the
number of compatible variables increases. However, when variables
are incompatible, it is unlike classical probability theory, and the
Hilbert space dimensionality remains constant as the number of incompatible
variables increases. In short, incompatibility\textendash a central
concept in quantum theory and its application to psychology \citep{plotnitsky2012niels,wang2015reintroducing}\textendash produces
simplification by rotating the basis to generate variables rather
than adding new dimensions. 

\subsection{The Hilbert space multi-dimensional program\protect\footnote{The word program here refers to the set of procedures that we formulated
to build HSM models. We are in the process of writing generalizable
computer codes to implement the conceptual program described here,
which will be published separately. At this point, we have created
computer codes for collections of $2\times2$ tables. The current
codes are written in Matlab and they are available at http://mypage.iu.edu/\textasciitilde{}jbusemey/quantum/HilbertSpaceModelPrograms.htm}}

An HSM model is built using the following programatic steps. All of
these steps are illustrated in the next section with a concrete application
to the artificial data set.

First, the researcher needs to determine which variables or attributes
are commutative, and which are not. Referring to our artificial data
set, we need to determine, for example, whether the attributes $A$
and $H$ commute or not. Recall that if they are compatible, then
they can be defined simultaneously, and sequence does not matter;
but if they are incompatible, they must be evaluated sequentially,
because one needs to change from one basis to another basis for evaluating
the sequence. One way to determine this is to observe whether or not
a pair of variables produce order effects. Alternatively, one can
statistically compare competing models with different hypothesized
compatibility relations.

Second, the dimension $N$ of the Hilbert space is determined. This
depends first of all on the assumed compatibility relations. Given
the compatibility relations, an HSM modeling procedure can begin with
the lowest possible dimension, and only increase the dimension as
required by model comparisons that favor a higher dimension. 

Third, a basis is selected for representing the coordinates of the
state vector $\left|\psi\right\rangle $ in terms of combinations
of compatible variables. Once a basis is chosen, the coordinates of
the state vector, represented by the $N\times1$ column matrix $\psi$,
can be estimated from the data. In general, each coordinate can be
complex, containing a magnitude and a phase. Therefore, if the dimension
equals $N,$ then the initial state requires $2\cdot N$ parameters.
However, the initial state must satisfy the unit length constraint
$\psi^{\dagger}\psi=1$, which constrains one magnitude. Also one
phase can be arbitrarily fixed without any effect on the choice probability.
In sum, only $2\cdot(N-1)$ parameters are estimated from the data.

Fourth, the projectors from unitary transformations are built, and
the latter are obtained by selecting a Hermitian operator used in
Equation \ref{eq:buildunitary}. In general, the Hermitian matrix
has $N$ diagonal entries that are real, and $N\cdot(N-1)/2$ off
diagonal entries, that can be complex. However, adding a constant
to all the diagonal entries has no effect on the choice probabilities,
and so one diagonal entry can be set to a fixed value. In sum, only
$\left(N^{2}-1\right)$ parameters are estimated for each Hermitian
matrix. 

Fifth, using quantum probability rules, the probability of a sequence
of measurements is computed. Using the predicted probabilities, the
model computes the log likelihood of the data given the model. The
parameters for the initial state and the Hermitian operators are estimated
from the data using maximum likelihood, and the model computes $G^{2}=-2\cdot loglikelihood$
statistics for model comparison.\footnote{Currently, we use a particle swarm method to estimate parameters in
order to avoid local minimum.} The number of model parameters is determined by the number of parameters
used to build the initial state vector plus the number of parameters
used to estimate the Hermitian operators used in Equation \ref{eq:buildunitary}.

Sixth, the fit of the model returns parameters for the initial state
that can be used to describe the probability distribution over a variable
as if it were measured alone (free of context of other variables),
and also the parameters of the unitary transformations that describe
the relations between incompatible variables.

Seventh, an HSM model allows many opportunities for very strong generalization
tests of the model. For example, if there are three variables, and
two of them are incompatible, then after estimating the model parameters
from an HSM model for a collection of $2\times2$ tables, the same
model and parameters can be used to make new predictions for new tables
that were not included in the original design, such as smaller $1-way$
tables or larger $3-way$ tables. 

\section{Application to the Artificial Data Set}

Step 1. Determine compatibility of variables. Psychologically, this
step determines whether two variables can be measured simultaneously
(compatible) or they have to be measured sequentially (incompatible).
Based on the order effects observed in Table \ref{tab:ArtificialData},
we infer that the pair of variables A,H were incompatible, as well
as the pair I,U. The design did not include manipulations of order
to test compatibility between variables A,I or H,U. In this case,
another way to empirically test compatibility is to compare model
fits that make compatibility vs. incompatibility assumptions about
these variables. Here for the purpose of illustration, we assumed
that they were compatible. 

Step 2. Define the Hilbert space. Assuming that A,I are compatible
means that we can define all of the events obtained from all of the
combination of values of these two variables: $(A=w\cap I=y)$, for
$(w=0,1)$ and $(y=0,1).$ Similarly, assuming that H,U are compatible
means that we can define all of the events formed by the all of the
combination of values of these two variables: $(H=x\cap U=z)$, for
$(x=0,1)$ and $(z=0,1).$ However, we cannot define combinations
for more variables because of the incompatibilities. The simplest
model is a model that assumes that each event is represented by only
one dimension, which produces a total of four dimensions. Therefore,
the minimum size of the Hilbert space was set to four dimensions,
and we started with this minimum.

Step 3. Define the initial state. We chose a basis that provided the
most meaningful parameters for the initial state. For this application,
we chose to use the basis defined by the combination of variables
A and I. Using this basis, the initial state $\left|\psi\right\rangle $
is represented by 
\begin{equation}
\left|\psi\right\rangle =\sum_{w,y}\psi_{wy}\cdot\left|A=w\cap I=y\right\rangle .\label{eq:Psyexample}
\end{equation}
The four coefficients in Equation \ref{eq:Psyexample} form a $4\times1$
column matrix 
\[
\psi=\left[\begin{array}{c}
\psi_{11}\\
\psi_{10}\\
\psi_{01}\\
\psi_{00}
\end{array}\right].
\]
For example, $\left|\psi_{10}\right|^{2}$ equals the probability
of yes to A and no to I when this pair of questions is asked in that
order. The parameters in $\psi$ are estimated from the data under
the constraint that $\psi^{\dagger}\psi=1$. In general, the 4 coefficients
can be complex valued, and so each coefficient contributes a magnitude
and a phase. However, the magnitudes must satisfy the unit length
constraint that $\psi^{\dagger}\psi=1$. Also, one phase for one coefficient
can be set to an arbitrary value without changing the final choice
probabilities. Therefore, only $4\times2-2=6$ free parameters are
required for the initial state. These parameters tell us what the
initial state of the psychological system (e.g., initial belief or
attitude towards attributes A and I) is before any measurement is
taken on the system, and can be used to compute the probability of
certain response to an attribute when it is measured alone. That is,
we can estimate more ``context free'' responses from the respondents\textendash free
from influences from measurement effects from the other attributes\textendash even
though we didn't collect such actual empirical data. 

Step 4. Define projectors and state transitions. Define $M_{n}=diag\left[\begin{array}{cc}
1 & 0\end{array}\right]$ , $M_{y}=diag\left[\begin{array}{cc}
0 & 1\end{array}\right]$, and $I_{2}=diag\left[\begin{array}{cc}
1 & 1\end{array}\right]$. The $4\times4$ matrix representation of the projector $P(A=y)$
is the Kronecker product $\left(M_{y}\otimes I_{2}\right)$, which
picks out the coordinates in $\psi$ that are associated with the
answer yes to attribute A. Likewise, the $4\times4$ matrix representation
of the projector $P(I=y)$ is the Kronecker product $\left(I_{2}\otimes M_{y}\right)$.
The $4\times4$ matrix representation of the projector $P(H=y)$ is
the Kronecker product $\left(U_{HA}M_{y}U_{HA}^{\dagger}\right)\otimes I_{2}$,
which requires the use of the unitary matrix $U_{HA}$ that transforms
coordinates from the A to the H basis. The $4\times4$ matrix representation
of the projector $P(U=y)$ is the Kronecker product $I_{2}\otimes\left(U_{UI}M_{y}U_{UI}^{\dagger}\right)$,
which requires the use of the unitary matrix $U_{UI}$ that transforms
coordinates from the I to the U basis.

The $2\times2$ matrix representations, $U_{HA}$ and $U_{UI}$, were
determined from Equation \ref{eq:Unitary} by selecting two $2\times2$
Hermitian matrices, $H_{HA}$ and $H_{UI}$. The parameters of each
of these Hermitian matrices were estimated from the data. Each $2\times2$
Hermitian matrix has four coefficients, two real diagonal values and
one complex off diagonal. However, one diagonal entry can be arbitrarily
fixed, and so only $4-1=3$ parameters are required for each $2\times2$
unitary matrix to produce a total of 6 parameters. These parameters
determine the rotation from the basis of one variable to the basis
of another variable. Psychologically, they tell us the relationship
between the variables or attributes being examined, and can reveal
the similarity between these variables, independent of the initial
state (i.e., step 3) of the person. In addition, based on the unitary
matrices, we can compute the probabilities the transition probabilities
between basis vectors. For example, assuming each answer is represented
by a single dimension, we can compute the transition probability from
a response to one attribute (e.g., say ``yes'' to the attribute
of attractiveness) to a response to another attribute (e.g., say ``no''
to the attribute of honesty). 

Step 5. Compute choice probabilities for each response sequence. The
choice probabilities for each sequence were computed by the product
of projectors corresponding to the sequence. For example, the probability
of YY for the AU table equals 
\begin{equation}
\left\Vert P(U=y)P(A=y)\left|\psi\right\rangle \right\Vert ^{2}=\left\Vert \left(I_{2}\otimes\left(U_{UI}M_{y}U_{UI}^{\dagger}\right)\right)\cdot\left(M_{y}\otimes I_{2}\right)\cdot\psi\right\Vert ^{2},
\end{equation}
the probability of NN for the HI table equals 
\begin{equation}
\left\Vert P(I=n)P(H=n)\left|\psi\right\rangle \right\Vert ^{2}=\left\Vert \left(I_{2}\otimes M_{n}\right)\cdot\left(\left(U_{HA}M_{n}U_{HA}^{\dagger}\right)\otimes I_{2}\right)\cdot\psi\right\Vert ^{2},
\end{equation}
the probability YN for the AH table equals 
\begin{equation}
\left\Vert P(H=n)P(A=y)\left|\psi\right\rangle \right\Vert ^{2}=\left\Vert \left(\left(U_{HA}M_{n}U_{HA}^{\dagger}\right)\otimes I_{2}\right)\cdot\left(M_{y}\otimes I_{2}\right)\cdot\psi\right\Vert ^{2},
\end{equation}
the probability NY for the HA table equals 
\begin{equation}
\left\Vert P(A=y)P(H=n)\left|\psi\right\rangle \right\Vert ^{2}=\left\Vert \left(M_{y}\otimes I_{2}\right)\cdot\left(\left(U_{HA}M_{n}U_{HA}^{\dagger}\right)\otimes I_{2}\right)\cdot\psi\right\Vert ^{2}.
\end{equation}

For the artificial data, assume that each $2\times2$ table was based
on 100 independent observations. The Hilbert space model has a total
of 12 free parameters, which is 3 less than the $4-way$ joint probability
model. Nevertheless, the Hilbert space model almost perfectly fits
all the relative frequencies in Table \ref{tab:ArtificialData}, and
the $G_{diff}^{2}=G_{H}^{2}-G_{Sat}^{2}=7.81\times10^{-6}$. 

Step 6. The 6 model parameters representing $\psi$, along with the
6 parameters representing $U_{HA}$ and $U_{IU}$, are presented in
the Appendix. Recall that the data were artificially generated for
illustration, and so the results are not to be taken seriously. However,
they help to show the application of a HSM model. We defer a more
detailed discussion of results until the next section, where we report
the results from fitting the model to real data from an experiment. 

\begin{table}
\caption{\label{tab:Predicted-one-way}Predicted probabilities of yes when
each variable is measured alone}

\centering{}%
\begin{tabular}{cc|c|c|c|}
\hline 
Attribute & A & H & I & U\tabularnewline
\hline 
\hline 
Probability & .4462 & .3691 & .3551 & .3843\tabularnewline
\hline 
\end{tabular}
\end{table}

The initial state, $\psi$, can be used to compute the probability
of responses to each attribute under the condition that the attribute
was measured alone (see Table \ref{tab:Predicted-one-way}). For example,
the predicted probability of answering Y (versus N) to the attribute
A when A is measured alone equals .4462 (probability of N equals $1-.4462=.5538$),
which is higher than the probability of answering yes to all the other
variables. Note that Table \ref{tab:Predicted-one-way} is not equal
to the marginal probability of a $2\times2$ or higher order table
involving pairs of incompatible variables.

The variables for a compatible pair, such as AI, are logically independent,
and the HSM model predicts the $2-way$ joint distributions for each
of the compatible pairs. (In this artificial case, the results are
almost perfectly predicted.) The variables for incompatible pairs,
such as AH and UI, are logically dependent, and so they do not provide
a joint distribution, and instead they produce a probability for a
sequence. 

The unitary matrices, $U_{AH}$ and $U_{UI}$, can be used to describe
the transitions between sequential measurements. The squared magnitudes
of the entries in the unitary matrices describe the probability of
transiting from a basis vector representing a column attribute (e.g.,
$\left|A=w\right\rangle \otimes\left|I=y\right\rangle $) to a basis
vector representing a row attribute (e.g., $\left|H=x\right\rangle \otimes\left|I=y\right\rangle $).
Table \ref{tab:Transitions} presents the transition probabilities
for the two incompatible pairs, AH, and IU. For example, the probability
of transiting to a positive answer for H from a positive answer to
A equals .$7738$, which is lower than the probability, .$8454$,
of transiting from a positive answer to I to a positive answer to
U. In other words, the variables I,U are more similar to each other
than the variables A,H. 
\begin{table}
\caption{\label{tab:Transitions}Transition matrices between basis vector for
pairs of incompatible attributes}

\centering{}%
\begin{tabular}{ccc}
 & $\left|A=0\right\rangle $ & $\left|A=1\right\rangle $\tabularnewline
\hline 
$\left|H=0\right\rangle $ & .7739 & .2261\tabularnewline
\hline 
$\left|H=1\right\rangle $ & .2261 & .7739\tabularnewline
\hline 
\end{tabular}%
\begin{tabular}{ccc}
 & $\left|I=0\right\rangle $ & $\left|I=1\right\rangle $\tabularnewline
\hline 
$\left|U=0\right\rangle $ & .8454 & .1546\tabularnewline
\hline 
$\left|U=1\right\rangle $ & .1546 & .8454\tabularnewline
\hline 
\end{tabular}
\end{table}

Step 7. The transition matrices produced by unitary matrices are always
symmetric. This is because each entry in the unitary matrix contains
the inner product between vectors from different bases, e.g. $\left\langle H|A\right\rangle $,
and the squared magnitude is the same in both directions, i.e., $\left|\left\langle H|A\right\rangle \right|^{2}=\left|\left\langle A|H\right\rangle \right|^{2}$.
Given the assumption in step 2 that events $(A=w\cap I=y),$ for all
$w,y,$ are represented by one dimensional subspaces (i.e., rays spanned
by basis vectors $\left|A=w\right\rangle \otimes\left|I=y\right\rangle $),
this implies symmetry in the conditional probabilities, i.e., $p(H=x|A=w)=p(A=w|H=x)$
and $p(I=x|A=w)=p(A=w|H=x)$. This is a very strong and empirically
testable property of this simple quantum model \citep{boyer2016testing}.
However, this symmetry does not hold generally; if events are represented
by multi-dimensional projectors instead of rays, then the conditionals
can be asymmetric (see, e.g., \citealt{PothosBusemeyerSim}). 

\section{An Empirical Application}

This section applies HSM modeling to a real experiment that was designed
in a manner similar to the artificial example. A total of 184 participants
made judgments on four attributes of anti-smoking public service announcements
(PSAs). They were asked to judge how Persuasive (P), Believable (B),
Informative (I), and Likable (L) they perceived various PSAs to be.
The PSAs were in the form of a single static visual image with a title.
Each person judged 16 different PSA's: One stimulus type included
8 examples warning about smoking causing death (Death PSAs), and the
other stimulus type included 8 PSA's warning about smoking causing
health harm (Harm PSAs). Each participant judged each PSA under 12
contexts: 6 combinations of two attributes with the attributes presented
in two different orders. For example, one context was PI, where the
participants answered the question of whether the PSA was Persuasive
and Informative by choosing either YY, YN, NY or NN (where for example
YN means Yes to Persuasive and No to informative). Thus each person
provided responses to $16$ (PSA's) $\times$$12$ (contexts) $=192$
questions, which were presented in a randomized order across participants.
Altogether, this produced a total of $184$ participants $\times$$192$
judgments per person $=35,328$ observations. 

The aggregate results, presented separately for each stimulus type,
but pooled across participants and order, are presented in Table \ref{tab:Pooledresults}
(later we present analyses at the individual level that include order.)
For example, when the Death PSA was presented, the relative frequency
of Y to Persuasive and N to Likable was .201, and the corresponding
result for the Harm PSA was .176. Each $2\times2$ table for a pair
of attributes is based on $184\cdot16=2944$ observations. However,
this table of pooled results ignores order effects and important individual
differences, and so the subsequent analyses were conducted at the
individual level of analysis.\footnote{Rather than conducting individual level analyses, we could formulate
a hierarchical Bayesian model that introduces assumptions about the
distribution of individual differences and priors on these hyper parameters.
At this early stage, we do not think this is a good place to start
for comparing complex models such as the $4-way$ joint probability
model (184 participants with $15\cdot2=30$ parameters for each participant)
because of lack empirical support for specific parametric distributions
of individual differences and lack of informative priors on the hyper-parameters
for these complex models. We did not want to confound our test of
core models (classical versus quantum) with arbitrary assumptions
about individual difference distributions and hyper priors. }

\begin{table}
\caption{\label{tab:Pooledresults}Observed Relative frequencies of pairs of
answers for 6 different pairs of attributes}

\centering{}%
\begin{tabular}{ccccc}
\multicolumn{5}{c}{Death PSA}\tabularnewline
\hline 
Attributes & YY & YN & NY & NN\tabularnewline
\hline 
PI & .529 & .166 & .072 & .232\tabularnewline
\hline 
PB & .612 & .092 & .074 & .223\tabularnewline
\hline 
PL & .501 & .201 & .064 & .235\tabularnewline
\hline 
IB & .539 & .074 & .128 & .259\tabularnewline
\hline 
IL & .441 & .181 & .127 & .251\tabularnewline
\hline 
BL & .495 & .188 & .086 & .232\tabularnewline
\hline 
\end{tabular}%
\begin{tabular}{ccccc}
\multicolumn{5}{c}{Harm PSA}\tabularnewline
\hline 
Attributes & YY & YN & NY & NN\tabularnewline
\hline 
PI & .438 & .134 & .049 & .379\tabularnewline
\hline 
PB & .459 & .099 & .069 & .374\tabularnewline
\hline 
PL & .378 & .176 & .083 & .362\tabularnewline
\hline 
IB & .419 & .078 & .109 & .394\tabularnewline
\hline 
IL & .324 & .169 & .124 & .383\tabularnewline
\hline 
BL & .356 & .184 & .102 & .359\tabularnewline
\hline 
\end{tabular}
\end{table}

\begin{table}
\caption{\label{tab:PredHSM-1}Probabilities predicted by the HSM model}

\centering{}%
\begin{tabular}{ccccc}
\multicolumn{5}{c}{Death PSA}\tabularnewline
\hline 
Attributes & YY & YN & NY & NN\tabularnewline
\hline 
PI & .544 & .155 & .065 & .236\tabularnewline
\hline 
PB & .610 & .064 & .055 & .271\tabularnewline
\hline 
PL & .507 & .192 & .069 & .232\tabularnewline
\hline 
IB & .539 & .071 & .132 & .258\tabularnewline
\hline 
IL & .441 & .142 & .124 & .293\tabularnewline
\hline 
BL & .493 & .178 & .083 & .246\tabularnewline
\hline 
\end{tabular}%
\begin{tabular}{ccccc}
\multicolumn{5}{c}{Harm PSA}\tabularnewline
\hline 
Attributes & YY & YN & NY & NN\tabularnewline
\hline 
PI & .444 & .122 & .061 & .373\tabularnewline
\hline 
PB & .486 & .064 & .056 & .394\tabularnewline
\hline 
PL & .386 & .180 & .079 & .355\tabularnewline
\hline 
IB & .428 & .077 & .109 & .386\tabularnewline
\hline 
IL & .356 & .142 & .124 & .378\tabularnewline
\hline 
BL & .361 & .176 & .104 & .359\tabularnewline
\hline 
\end{tabular}
\end{table}

\subsection{Test of the joint probability model}

\begin{figure}

\caption{\label{fig:Quantile---quantile}Quantile - quantile plot of the observed
versus predicted chi-square value}

\centering{}\includegraphics[scale=0.33]{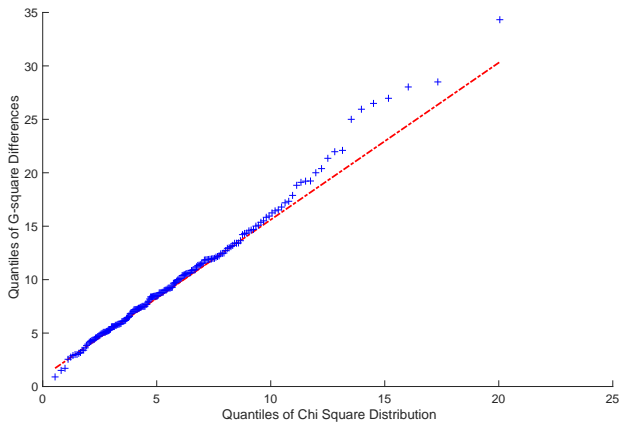}
\end{figure}

Each individual produced a table in the same form as Table \ref{tab:Pooledresults},
but with 16 observations per $2\times2$ table (192 observations in
total for both types of stimuli). Recall that the joint probability
model states that the 6 rows of $2\times2$ tables are produced by
a joint distribution, $\pi(P=w\cap B=x\cap I=y\cap L=z)$ where $w=0,1,$
$x=0,1$, $y=0,1$ and $z=0,1$, that has $16-1=15$ free parameters
per stimulus type or $30$ parameters altogether. The saturated model
requires $3$ parameters for each $2\times2$ table, producing a total
of $18$ parameters per stimulus type or $36$ parameters altogether.
Using maximum likelihood estimation for each person, we computed the
$G_{sat}^{2}$ and $G_{joint}^{2}$ for each person. A quantile-quantile
plot of the observed $G^{2}$ differences, $G_{diff}^{2}=G_{joint}^{2}-G_{sat}^{2}$
versus the $\chi^{2}$ predicted by the null hypothesis is shown in
Figure \ref{fig:Quantile---quantile}. As can be seen in Figure \ref{fig:Quantile---quantile},
the observed $G_{diff}^{2}$ exceeds the expected for large values
of the predicted chi-square. We computed a lack of fit from the null
chi-square distribution by comparing the observed versus expected
frequencies using categories defined by cutoffs {[}0, 5, 10, 35{]}.
The expected frequencies were $[84,77,23]$ but the observed frequencies
were $[48,75,61]$, and the difference is statistically significant
($\chi^{2}(2)=78.84$). We conclude that the $4-way$ joint probability
model systematically deviates from the observed results for a substantial
number of individuals. 

\subsection{Comparisons between Bayes net and HSM models}

Any Bayesian network model, based on the four random variables, P,B,I
and L, is a special case of the $4-way$ joint probability model,
which implies that there is also some systematic deviation from any
Bayes net type of model. However, there may also be systematic deviations
from a HSM model. Therefore, it is important to compare the fits of
Bayes net versus HSM models. Because they two types of models are
non-nested, we performed comparisons at the individual level using
the Bayesian information criterion $BIC_{Model}=G_{Model}^{2}+p\cdot ln(192)$,
where $G_{Model}^{2}=-2\cdot loglikelihood$, and $p=$number of model
parameters.

\subsection{Simple Bayes net model}

There are a large number of possible Bayes net type of models that
one can construct for this application. We chose the following model
because (a) it is simple and (b) it makes assumptions that match design
of the stimuli and responses to the stimuli for this experiment. We
note, however, that our conclusions are restricted to these particular
models, and there may be other Bayes net models that perform better
than the one here. 

For the Bayes net type of model, we assumed that the two attributes,
Informative (I) and Believable (B), are exogenous factors determined
by the type of PSA's. Therefore, each type of stimulus produced a
$2-way$ joint distribution with four joint probabilities $\pi(I=x\cap B=y|stimulus),$
$x=0,1,$ $y=0,1$, and there are $2$ types of stimuli. This produces
$(4-1)\cdot2=6$ parameters per stimulus type. Next we assumed that
the response to attributes Persuasive (P) and Likable (L) depended
on the stimulus attributes I and B, which was represented by the conditional
probabilities $\pi(P=w\cap L=z|I=x\cap B=y)$, for $w=0,1$ and $z=0,1$.
However, this model produces the same number ($15\cdot2=30$) of parameters
as the $4-way$ joint probability model. To simplify the model, we
assumed independence, so that $\pi(P=w\cap L=z|I=x\cap B=y)=\pi(P=w|I=x\cap B=y)\cdot\pi(L=z|I=x\cap B=y)$.
We also assumed that the two conditionals, $\pi(P=w|I=x\cap B=y)$
and $\pi(L=z|I=x\cap B=y)$, did not depend on the stimulus type.
Therefore, each of the two conditionals produces four parameters.
Altogether, this model entails $(4-1)\cdot2+\left(4\cdot2\right)=14$
parameters. 

\subsection{Simple HSM model}

The same simple HSM model used to fit the artificial data in the previous
section was applied to the real data from our experiment. First, we
assumed that the attributes, Believable (B) and Informative (I) are
compatible, which means we can think of these two attributes at the
same time and the order of measuring the two attributes does not matter.
This is consistent with the lack of effect of the order effects of
the two attributes in the aggregated data. Second, we assumed that
Persuasive (P) is a rotation of Believable, and Likable (L) is a rotation
of Informative. In other words, B,P were assumed to be incompatible
and so were I,L. This assumption was also consistent with order effects
found at the aggregate level for these variables. For simplicity,
we assumed that each joint event for a compatible pair, such as $(B=x\cap I=y)$,
is represented by a single dimension. We chose to represent the initial
state and projectors by the basis described by the B,I events $(B=x\cap I=y).$ 

To reduce the number of model parameters to a minimum, we restricted
the coordinates of the initial state to be a real valued $4\times1$
matrix $\psi$ with unit length $\psi^{\dagger}\psi=1$. The unitary
matrix for rotating between the incompatible basis vectors was constructed
using a single parameter as follows:
\begin{eqnarray}
H & = & i\cdot\theta\cdot\left[\begin{array}{cc}
0 & -1\\
1 & 0
\end{array}\right],\nonumber \\
U & = & exp\left(-i\cdot H\right)\nonumber \\
 & = & \left[\begin{array}{cc}
cos\left(\pi\cdot\theta\right) & -sin\left(\pi\cdot\theta\right)\\
sin\left(\pi\cdot\theta\right) & cos\left(\pi\cdot\theta\right)
\end{array}\right].
\end{eqnarray}
Parameters, $\theta_{PB}$ and $\theta_{LI}$ , were used to define
rotation matrices $U_{PB}$ for the P,B incompatible pair and $U_{LI}$
for the I,L incompatible pair. 

To account for the effect of type of stimulus, we allowed the initial
state vector to vary across stimuli, $\psi_{Death}$ and $\psi_{harm}$.
However, according to the HSM model, the transitions between basis
states for incompatible variables B and P, as well as the transitions
between basis states for incompatible variables I and U, should only
depend the unitary transformation $U,$ and the latter depends only
on the variables and is independent of the stimulus. We tested this
interesting prediction from HSM model by comparing a model that allowed
$\theta_{PB},\theta_{LI}$ to change across stimuli with a model that
constrained these to be the same across stimuli. The constrained HSM
model produced a total of $\left(3\cdot2\right)+2=8$ parameters. 

In sum, the HSM model starts with the $4$-dimensional BI basis, which
provides the coordinates that define the initial distribution $\psi$.
The coordinates of $\psi$ are then used to compute the $2-way$ joint
distribution for the BI table. The distributions for all of the other
$2-way$ tables are generated by rotating the basis of the $4$ dimensional
space using the unitary matrices $U_{LI}$ and $U_{BP}$. The new
basis produced by rotation provides coordinates that are then used
to compute the response probabilities for another table. 

\subsection{Results of model comparisons}

\begin{figure}

\caption{\label{fig:BIC}Frequency of BIC differences (BIC Bayes - BIC HSM)}

\begin{centering}
\includegraphics[scale=0.33]{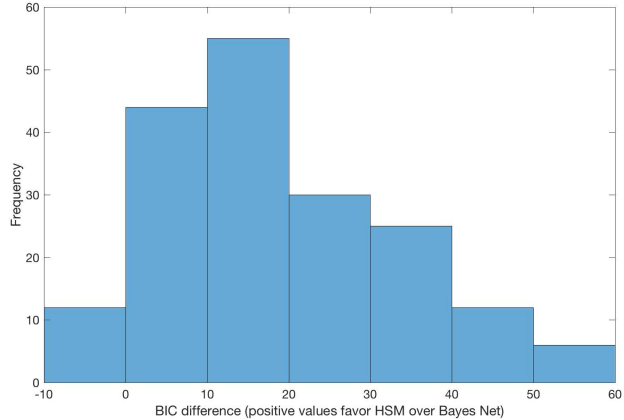}
\par\end{centering}
\end{figure}

Maximum likelihood estimates and $G^{2}$ statistics were computed
by fitting each model the model to the 192 observations separately
for each participant. When comparing the $4-way$ joint probability
model (30 parameters) to the constrained HSM model (8 parameter model),
all 184 participants produced a BIC difference favoring the constrained
HSM model. More interestingly, when comparing the Bayes net model
(14 parameters) to the constrained HSM model (8 parameters), 148 out
of 184 participants produced BIC differences favoring the constrained
HSM model. The distribution of BIC differences ($BIC_{diff}=BIC_{Bayes}-BIC_{HSM})$
is presented in Figure \ref{fig:BIC}. 

Finally, we tested the prediction of the HSM model that the rotation
parameters are invariant across stimuli by comparing the $BIC_{constrained}$
to $BIC_{unconstrained}$ versions of the HSM model. In agreement
with the prediction, the $BIC$ difference favored the constrained
model for 154 out of 184 participants.

The predictions generated by the constrained HSM model, pooled across
participants and presentation order are presented in Table \ref{tab:PredHSM-1}.
As can be seen in the table, the constrained HSM model does a very
good job of predicting the pooled results. The most important errors
occur for the incompatible variables, where we constrained the model
to use the same parameters across stimulus types.

\subsection{Interpretation of parameters}

\begin{figure}
\caption{\label{fig:PsyFreq}Relative frequency distribution for the probabilities
of a ``yes'' answer to each of the I, B, L, and P attribute when
measured alone (left panels, death PSAs; right panels, harm appeal
PSAs)}

\centering{}\includegraphics[scale=0.5]{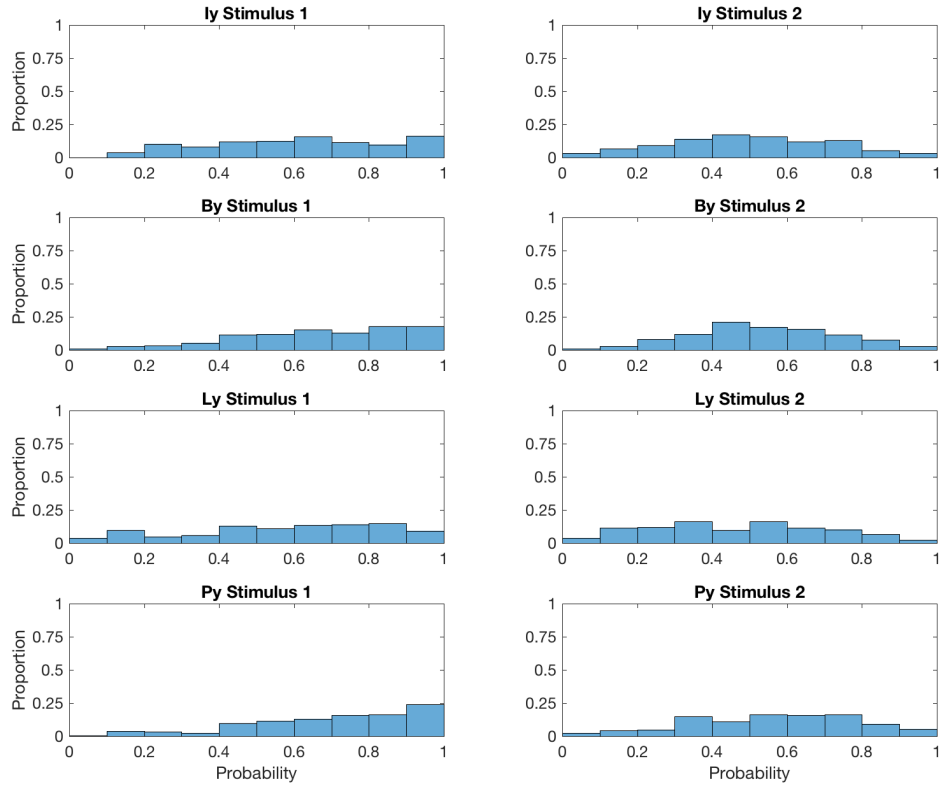}
\end{figure}

The HSM model provides two sets of model parameters for each participant.
One set, which is based on the initial state $\psi$, describes the
probabilities of responding ``yes'' to each variable when the variable
is measured alone (free from context effects of other attributes).
Figure \ref{fig:PsyFreq} presents the relative frequency distribution
of these response probabilities for each type of stimulus. For example,
the bottom left panel shows the relative frequencies for ``yes''
responses to P attribute with the death appeal PSAs, and the right
lower panel shows the results for the harm appeal PSAs. As can be
seen in the figure, the probabilities are widely spread out among
participants, but the probability of answering ``yes'' was generally
higher for the death appeal PSAs. Similarly, we can compare the parameter
distributions for the other three attributes between the two types
of PSAs with different appeals (see Figure 3). In general, participants
responded more positively towards death appeal PSAs on all the four
attributes, but clearly more so for the attributes of believable and
persuasive. 

\begin{figure}

\caption{\label{MuPlot}Relative frequency of transition probability from yes
in one basis to yes in another}

\begin{centering}
\includegraphics[scale=0.33]{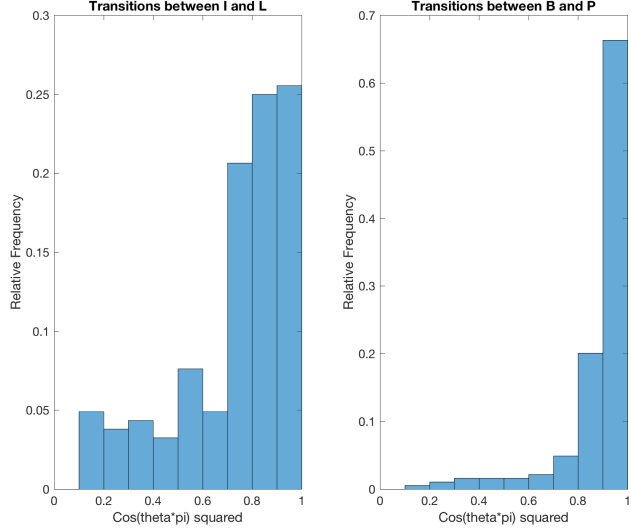}
\par\end{centering}
\end{figure}

The second set is based on the parameters $\theta{}_{PB},\theta_{LI}$
used for the rotation matrices for the two incompatible variables
(recall that these are the same for the two types of stimulus). The
squared magnitude of the coefficients within the unitary rotation
matrices describe the probability of transiting from one basis to
another, that is, transitioning from basis vectors for I to basis
vectors for L and transitioning from basis vectors for B to basis
vectors for P. Figure \ref{MuPlot} presents the relative frequency
of $cos(\theta\cdot\pi)^{2}$, which describes the probability of
transiting from a ``yes'' to one variable to a ``yes'' to another
variable that is incompatible with the former variable. The panel
on the left presents the distribution for $\theta_{IL}$ and the distribution
on the right is for $\theta_{PB}$. As can be seen in Figure 4, the
parameter for each pair of attributes is located at a high value on
average, indicating that the two attributes are quite similar to each
other. Interestingly, however, the similarity between P and B tends
to be higher across all participants than that for L and I; in addition,
there are larger individual differences for the L and I transitions
since the parameter distribution is more widely distributed compared
to that for the P and B transitions (see Figure 4).

\section{Summary and Extensions}

In this article, we presented the general procedures for building
HSM models based on quantum probability theory. These models provide
a simple and low dimensional vector space representation of collections
of contingency tables formed from measurement of subsets of $p$ variables.
HSM models are needed when responses to questions about a variable
depend on the context formed by the other variables present in the
subset. HSM models provide tools for modeling context effects, and
the model parameters provide two psychologically meaningful and useful
interpretations of these effects. First of all, the state vector of
an HSM model provides an estimation of the respondents\textquoteright{}
initial response tendencies to each of the $p$ variables in a context
free manner, that is, as if a variable was measured in isolation.
Second, the measurement operators describe the inter-relations between
the $p$ measurements, independent of the initial response tendencies.
Furthermore, once the variables being measured have been mapped into
the Hilbert space by an HSM model, the parameters of the model can
be used to make new predictions for new contexts not included in the
original design.

HSM models provide new contributions to the current set of probabilistic
and statistical tools for contingency table analysis. Loglinear/categorical
data models only apply to a single table containing all $p$ variables,
whereas the HSM models can be applied to multiple tables containing
different subsets of the $p$ variables. Bayesian network models can
also be applied to collections of tables; however, they assume the
existence of a complete $p-way$ joint distribution, and it is often
the case that no complete $p-way$ joint distribution exists. HSM
models can be applied to collections of tables even when no $p-way$
joint distribution exists to reproduce the collection.

In addition to presenting the general procedures for constructing
HSM models, we presented an artificial data example and a real data
example. The artificial example was designed to illustrate (a) violations
of consistency requirements of the $p-way$ joint distribution model,
(b) a non-parametric statistical test of a $p-way$ joint distribution
for a collection of tables, and (c) illustrate the application of
an HSM model to a concrete example. The real data example (a) presented
the results of a new experiment investigating evaluations of health
messages, (b) reported significant deviations from the $4-way$ joint
distribution, and (c) compared the fit of a simple HSM model to a
simple Bayes net model using Bayesian information criteria. We conclude
from these analyses that HSM models are empirically viable for modeling
collections of contingency tables.

Besides those considered here, many other applications of HSM models
are possible. For example, past research in consumer behavior has
shown that measurements of preferences for different sets of consumer
products are context dependent \citep{huber1982adding}, and HSM models
could be used to analyze these context effects. As another example,
the HSM models can be useful for analyzing survey data from multiple
sources such as different family members or different cross-cultural
groups \citep{de2012clusterwise}. Dynamic extensions of HSM models
can be used to model changes in measurements across longitudinal or
multiple stage surveys when different subsets of measurements are
used across stages \citep{mcardle2009modeling}. In sum, HSM models
can be applied to complex data collected from a large number of different
sources and contexts found in the social and behavioral sciences. 

\bibliographystyle{plainnat}
\bibliography{Busemeyer.bib}

\section{Appendix}

\subsection{Matrix exponential function}

Suppose $H$ is the matrix representation of a Hermitian operator.
Then we can decompose $H$ into its orthonormal eigenvector matrix
$V$ and its real eigenvalue diagonal matrix $\Lambda$ as follows:
$H=V\cdot\Lambda\cdot V^{\dagger}$. The matrix exponential of $H$
is defined as 
\begin{align*}
exp(H) & =V\cdot exp(\Lambda)\cdot V^{\dagger},\\
exp(\Lambda) & =diag\left[\begin{array}{ccccc}
e^{\lambda_{1}} & \cdots & e^{\lambda_{j}} & \cdots & e^{\lambda_{N}}\end{array}\right].
\end{align*}

\subsection{Kronecker product}

Suppose $P$ is a $m\times n$ matrix and $Q$ is a $r\times s$ matrix.
Then the Kronecker product is a $\left(m\cdot r\right)\times\left(n\cdot s\right)$
matrix defined by 
\[
P\otimes Q=\left[\begin{array}{ccccc}
p_{11}\cdot Q &  & \vdots &  & p_{1n}\cdot Q\\
\vdots &  & \vdots &  & \vdots\\
\vdots & \cdots & p_{ij}\cdot Q & \cdots & \vdots\\
\vdots &  & \vdots &  & \vdots\\
p_{m1}\cdot Q &  & \vdots &  & p_{mn}\cdot Q
\end{array}\right].
\]
For example, 
\[
\left[\begin{array}{ccc}
2 & 3 & 4\\
3 & 6 & -2\\
4 & -2 & 5
\end{array}\right]\otimes\left[\begin{array}{cc}
1 & 0\\
0 & 1
\end{array}\right]=\left[\begin{array}{cccccc}
2 & 0 & 3 & 0 & 4 & 0\\
0 & 2 & 0 & 3 & 0 & 4\\
3 & 0 & 6 & 0 & -2 & 0\\
0 & 3 & 0 & 6 & 0 & -2\\
4 & 0 & -2 & 0 & 5 & 0\\
0 & 4 & 0 & -2 & 0 & 5
\end{array}\right]
\]
The Kronecker product satisfies the following property (assuming the
column dimension of $P$ matches the row dimension of $U,$ and likewise
for $Q$ and $T):$ 
\[
\left(P\otimes Q\right)\cdot\left(U\otimes T\right)=\left(P\cdot U\right)\otimes\left(Q\cdot T\right).
\]

\subsection{Parameters used to fit artificial data}

\[
\psi=\left[\begin{array}{c}
.5203\cdot e^{0}.\\
.4189\cdot e^{i\cdot2.2920}\\
.2904\cdot e^{i\cdot0.9383}\\
.6852\cdot e^{i\cdot0400}
\end{array}\right]
\]
Note that $.5203=\sqrt{1-\left(.6852^{2}+.2904^{2}+.4189^{2}\right)}$. 

\[
H_{HA}=\left[\begin{array}{cc}
-.5911 & -.5037\cdot e^{i\cdot.8862}\\
-.5037\cdot e^{-i\cdot.8862} & 0
\end{array}\right]
\]
Table \ref{tab:Transitions} is obtained by squaring the magnitudes
of the entries.

\[
H_{UI}=\left[\begin{array}{cc}
-1.2405 & -.4334\cdot e^{i\cdot1.2976}\\
-.4335\cdot e^{-i\cdot1.2976} & 0
\end{array}\right]
\]
Table \ref{tab:Transitions} is obtained by squaring the magnitudes
of the entries. 
\end{document}